# Direct visualization of topological transitions and higher-order topological states in photonic metasurfaces

Anton Vakulenko[1], Svetlana Kiriushechkina[1], Mengyao Li[1,2], Dmitriy Zhirihin[1], Xiang Ni[1,2,3], Sriram Guddala[1], Dmitriy Korobkin[3], Andrea Alù[3,2,1], and Alexander B. Khanikaev[1,2]

[1]Department of Electrical Engineering, Grove School of Engineering, City College of the City University of New York, 140th Street and Convent Avenue, New York, NY 10031, USA.

[2]Physics Program, Graduate Center of the City University of New York, New York, NY 10016, USA.

[3]Photonics Initiative, Advanced Science Research Center, City University of New York, New York, NY 10031, USA

**Abstract:** Topological photonic systems represent a new class of optical materials supporting boundary modes with unique properties, not found in conventional photonics. While the early research on topological photonics has focused on edge and surface modes in 2D and 3D systems, respectively, recently higher-order topological insulators (HOTIs) supporting lower-dimensional boundary states have been introduced. In this work we design and experimentally realize a photonic kagome metasurface exhibiting a Wannier-type higher-order topological phase. We demonstrate and visualize the emergence of a topological transition and opening of a Dirac cone by directly exciting the bulk modes of the HOTI metasurface via solid-state immersion spectroscopy. The open nature of the metasurface is then utilized to directly image topological boundary states. We show that, while the domain walls host 1D edge states, their bending induces 0D higher-order topological modes confined to the corners. The demonstrated metasurface hosting topological boundary modes of different dimensionality paves the way to a new generation of universal and resilient optical devices which can controllably scatter, trap and guide optical fields in a robust way.

**Introduction**. Topological photonics[1,2,3] offers new approaches to control and manipulate electromagnetic radiation by exploiting synthetic gauge fields to imitate exotic quantum solid-state phenomena[2,3,4,5,6,7,8,9,10,11,12,13,14,15,16,17]. In the past decade, several approaches to confine and guide electromagnetic waves in a robust way have been demonstrated in a variety of platforms, from arrays of silicon ring resonators[6,10,18] and coupled waveguide arrays[9,19,20], to photonic crystals[5,11,13,21,22,23] and metamaterials[8,14,17,24,25,26]. The early topological photonic materials were focused on trapping topological modes along interfaces - domain walls and boundaries – with their dimensionality being one lower than the material itself, and used them for a variety of applications from robust guiding, delay lines, in tunable devices, to lasing[24,25,26,27,28,29,30,31]. The more recent concept of higher-order topological insulators (HOTIs) has opened a wider range of topological modes confined to even lower dimensional boundaries[32,33,34,35]. Following the early theoretical predictions of photonic HOTIs, a number of experimental realizations have been reported in mechanics[36] and acoustics[37,38,39,40], and in electromagnetics, both in the microwave[41,42,43,44] and optical spectral domains [20,45,46,47]. The significant interest in HOTIs is not surprising since, in addition to the purely scientific curiosity, the ability to confine modes in a robust way in any number of dimensions can be of fundamental importance for practical applications. For 2D topological photonic systems, where conventional 1D guided topological modes have been proposed for robust transport of signals, lower-dimensional higher-order topological 0D states may be used as resilient resonators. The possibility to dynamically couple modes in this dimensional topological hierarchy farther expands the topological toolkit by offering robust trapping, release and guiding of electromagnetic energy over the same platform.

Research on photonic HOTIs (PHOTIs) has been very fruitful at microwaves, with demonstrations of 2D systems with multipolar (quadrupole[41]) and dipolar bulk polarizations supporting topological corner states. In the optical domain, a pioneering work by Noh[20] has demonstrated the possibility of confining modes at corners of 2D waveguide arrays. A quadrupolar PHOTI was implemented in an array of coupled ring resonators[45], and more recently the trapping of light at the nanocavity formed by the 90-deg-angled rim of a two-dimensional photonic crystal was demonstrated[47]. Moving these concepts to ultrathin metasurfaces offers even more possibilities to control electromagnetic radiation, in particular by coupling these topological states to the radiation continuum, with the opportunity of controlling in new ways the scattering of light. It has been recently shown that topological transitions in metasurfaces can be used to control the radiative properties of resonant modes[48, 49], and the first topological grating based on an array of topological domain walls has been demonstrated.

In this work, we experimentally demonstrate a new type of PHOTI metasurface and we directly visualize the changes in its band structure due to topological transition. The open character of the structure enables direct imaging of corner and edge states supported at the boundaries of the topologically nontrivial metasurface domains.

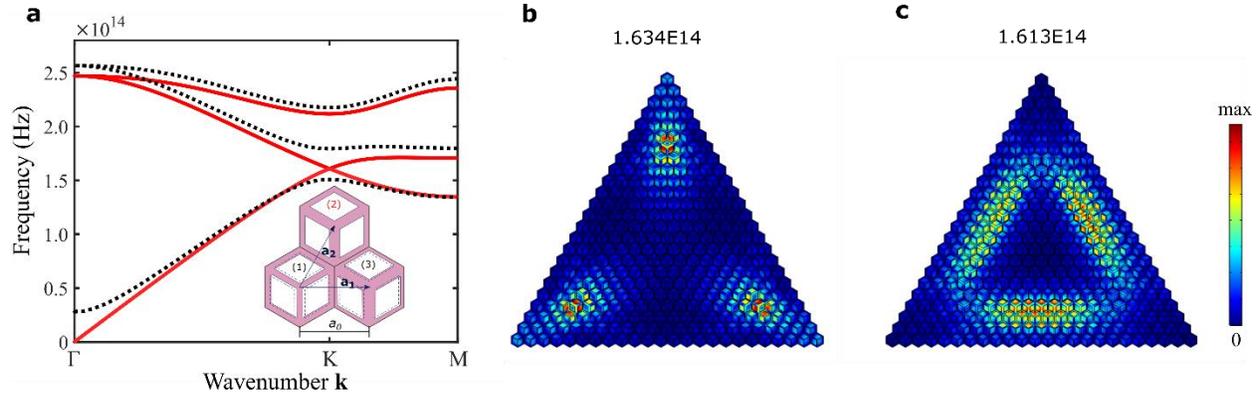

**Fig. 1| Band structure and boundary modes of a topological 2D kagome crystal. a** Schematics of the 2D Kagome lattice (inset) with diamond shaped holes, and its band structure calculated by finite-element method (FEM) for different scenarios. Red lines correspond to the unperturbed lattice (inset, geometry 2) and black dashed lines represent shrunken (inset, geometry 1) and expanded (inset, geometry 3) lattices (identical band structures). **b**, **c** Field profiles for corner and edge eigenmodes, respectively, calculated for a finite expanded Kagome lattice surrounded by the shrunken lattice. $a_0$ is the lattice constant, $\mathbf{a_1}$ and $\mathbf{a_2}$ are lattice vectors in real space. Parameters used: $a_0 = 520$ nm, the long size the air hole is $0.71a_0$ the short one is $0.41a_0$. The structure is expanded or shrunken by 10%. The numbers on top of **b** and **c** are eigenfrequencies of the boundary modes.

**Theoretical analysis.** The unit cell of the proposed PHOTI is shown in the inset of Fig. 1a (geometry 2), and it represents a photonic crystal with kagome lattice formed by trimers of diamond shaped holes (filled with air $\epsilon = 1$) in high-index silicon ($\epsilon = 12$) matrix. The triangular symmetry of the array ensures that the kagome lattice possesses a degeneracy point (Dirac point) at the K-point in the Brillouin zone, and the choice of shape and size of the diamond holes ensures that the Dirac point is embedded within a sufficiently wide bandgap, which can be seen for the case of TE polarization in Fig.1a (red lines). This property ensures the presence of a complete topological bandgap after the topological transition takes place. Kagome systems in electronic[35] and acoustic systems[50] have been shown to support a topological transition, accompanied by bandgap opening, when the symmetry of the system is reduced by dimerizing the lattice, i.e., by inducing interactions within the unit cell formed by the trimer and between different unit cells unequal. It is possible to dimerize the system in two ways: by making the interactions within or outside the unit cell stronger, yielding two topologically distinct phases characterized by different topological bulk polarization, (1/3,1/3) and (0,0), respectively[37]. The bulk polarization characterizes average displacement of the wavefunction with respect to the unit cell center (along basis vectors) as the wavenumber shifts between two time-reversal invariant points. It also establishes the bulk-interface correspondence [32, 51] ensuring that proper cuts of the topological system, with nonzero polarization, or certain boundaries between distinct topological domains, must host topological boundary modes - 1D edge modes or higher-order 0D corner states[37].

In order to induce a topological transition in the proposed photonic Kagome structure, we reduce the symmetry of the lattice by bringing diamond-shaped holes closer ("shrink" the trimer, as in geometry 1 of Fig. 1**a** inset) or by moving them farther apart ("expanding" the trimer, as in geometry 3 of Fig.1**a** inset). While the band structures for the shrunk and expanded cases of the 2D photonic crystal, shown in Fig. 1**a** by dashed lines, are identical, the two cases can be discriminated by the direct evaluation of the Wannier center and of the topological bulk polarization from the numerically calculated eigenmodes of the crystal. Our first-principle calculations show that the Wannier center shifts by 1/3 as the wavenumber changes from the Γ-point to the K-point for the expanded case, but not for the shrunken case, which confirms distinct topological phases of the two crystals. The numerical calculations for the case of a finite triangular-shaped topological (expanded) domain surrounded by the trivial (shrunken) domain further confirm that both edge (Fig. 1**b**) and higher-order boundary modes (Fig. 1**c**) emerge at the domain walls and at their corners, respectively.

In order to confirm the presence of topological boundary modes for our open silicon metasurface, even though the modes may be compatible with radiation in terms of their transverse momentum, and hence may be allowed to leak their energy into free space, we performed first-principle modelling of a 220nm-thick structure with SiO$_2$ substrate $\epsilon_{\text{sub}} = 2.1$ and air superstrate $\epsilon_{\text{sup}} = 1$, for the cases of perturbed (shrunken or expanded) and unperturbed (ideal Kagome) lattices. The band structures, shown in Fig. 2**a**, again manifest a Dirac-like dispersion (red solid lines) and gapped bands for the unperturbed and perturbed scenarios, respectively. Note, however, that in this case the modes arise also above the light cones for some values of the Bloch wave-vector **k**, implying that they become leaky in that region of the band diagram. Despite their radiative nature, the Wannier-center calculations again confirm that shrunken and expanded metasurfaces are characterized by topological polarizations of (0,0) and (1/3,1/3), respectively[37]. Although our system is non-Hermitian due to its open nature, which can yield new interesting aspects of topology,[52, 53, 54, 55] in our case, since the bulk bands predominantly appear below the light cones, the radiation leakage is small enough to not affect the original topological nature of the 2D bulk modes. On the other hand, the unavoidable leakage of the boundary modes, both 1D edge and 0D corner states, provides a unique opportunity to excite and image these states directly from the far-field[48].

Next, we performed first-principle calculations of the finite triangular-shaped topological (expanded) metasurface laterally surrounded by the trivial (shrunken) domain with same parameters. As shown in Fig. 2**b,c**, the simulation results for plane-wave excitation (conducted with CST Microwave Studio for the angle of incidence of 70 deg.) clearly confirm the presence of two types of boundary modes. These results demonstrate that, due to the open nature of the metasurface, the topological boundary modes couple to the radiation continuum, and thus can be excited from and can radiate into the far-field. The first type of excited modes clearly represents 0D higher-order states localized at the corners of the topological crystal (Fig.2**b**), while the second type consists of edge states, localized at the 1D domain walls between trivial and topological metasurfaces (Fig.2**c**).

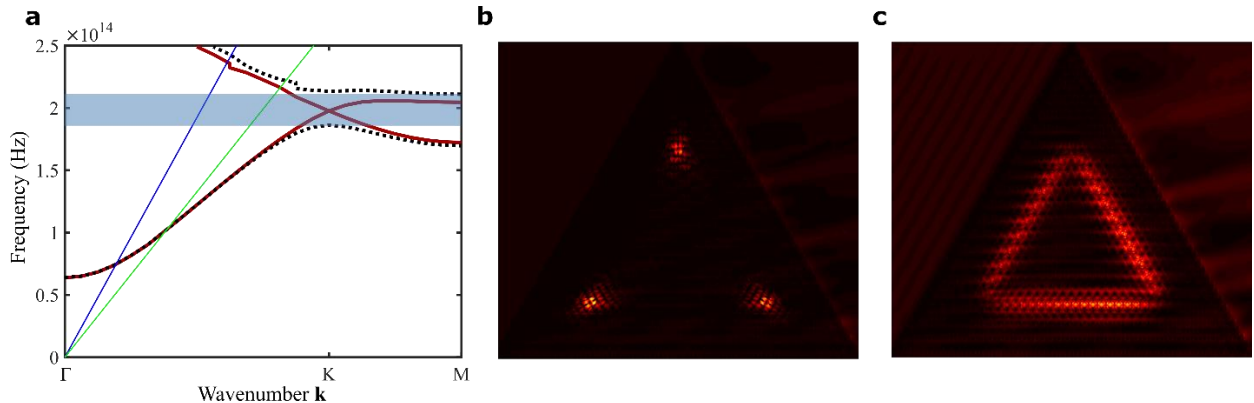

**Fig. 2| Simulation results of corner and edge states for the 3D Kagome cladded silicon-on-insulator (SOS) metasurface**. **a** 3D Kagome band simulation, unperturbed (red) and expanded (black dotted). Blue band shows the bandgap region, blue and green lines indicate light lines of air and substrate (SiO$_2$) respectively. **b** and **c** show corner and edge states, respectively, as obtained from plane wave excitation in CST Microwave studio simulation. Geometric parameters are as in Fig. 1.

**Experimental results.** To experimentally confirm opening of the complete photonic band gap due to symmetry reduction, we fabricated large (300$\mu$m × 300$\mu$m) samples of metasurfaces (with over 300,000 unit cells) to facilitate the band structure recovery. Since the Dirac cone and the subsequent bandgap opening take place below the air light line, we exploited solid immersion spectroscopy, by coupling light to the modes of the structure through a high-index ($n=2.45$) ZnSe hemispherical lens brought in immediate proximity with the sample. Subsequently, attenuated total reflection (ATR) measurements through the lens with varying angle of incidence ($\theta_{inc}$ from 35° to 65°) and azimuthal angle ($\phi$ from $-5°$ to 5°) enabled scanning of evanescent modes supported by the metasurface with the wavenumbers in the proximity of the K-point, $\mathbf{k}_K = (\frac{4\pi}{3a_0}, 0)$, in two directions. The experimentally recovered surfaces of the bands for both ideal (unperturbed) kagome and expanded (perturbed) metasurfaces are plotted in Fig. 3, alongside with theoretical results. The experimental data clearly reveal the presence of a point degeneracy and of a Dirac cone for the unperturbed metasurface, as well as lifting of the degeneracy and opening of the band gap for the perturbed structure. This is, to our knowledge, the first direct visualization of a Dirac cone and of topological transitions in optics. Detailed description of the band structure reconstruction and a schematic of the corresponding setup are given in Methods.

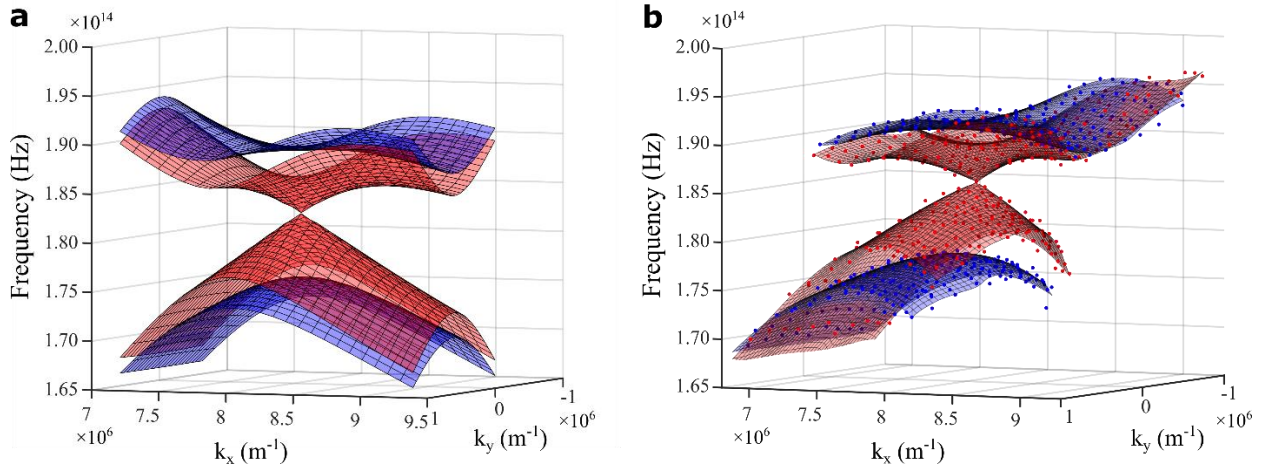

**Fig. 3| Theoretical and experimental Dirac cone and topological bandgap for the kagome metasurfaces**. Band surfaces near K point (**a**) from FEM simulations and (**b**) reconstructed form the solid immersion experiment. In both panels, red surfaces reveal Dirac cones for the unperturbed kagome metasurface, and the blue bands reveal a bandgap for perturbed kagome lattices. Red and blue dots are the extracted experimental data. The surfaces in (**b**) are obtained by fitting experimental points with a Dirac cone function with (5$^{th}$ degree) polynomial corrections.

In order to experimentally prove the existence of boundary modes, smaller-sized samples of our triangular shaped topological (expanded) crystals surrounded by larger-sized trivial (shrunken) metasurface were fabricated by e-beam lithography (see Fig.4**a** for the enlarged SEM picture near one of the corners of the structure, and the fabrication process is described in Methods). The samples were then tested by directly focusing light of a fixed wavelength from a high-intensity supercontinuum light-source (SuperK Extreme) with tunable high-resolution bandpass filters (LLTF Contrast). Light was incident at a large angle (~70°) to facilitate dark-field imaging by a reflective (Schwarzschild) high-NA objective and near-IR camera (XEVA InGaAs) at a direction perpendicular to the metasurface. The localized character of the boundary modes enabled direct coupling to focused incident light and subsequent radiation of the trapped light into free space with a broad range of wave-vectors, thus facilitating direct dark-field imaging. A detailed description of the imaging setup is given in Methods.

The experimentally mapped spatial distribution of the field as a function of wavelength allows us to extract the density of states for the topological boundary modes, shown in Fig. 4**b**. Figure 4**b** reveals two characteristic peaks when specific (spatial) filter-functions are applied to the experimental field profiles. To characterize the density of corner states, we pick fields in the vicinity of three corners of the topological triangle (square with side of $4a_0$), while for the edge states the filter function represents three strips of the same width ($4a_0$) with domain walls in the center schematically shown in the inset to Fig. 4**b.** The inspection of the filed profiles at the peak maxima, shown in Fig. 4**c,d**, confirms the mode localization at the corners and edges of the triangle, thus evidencing their topological origin.

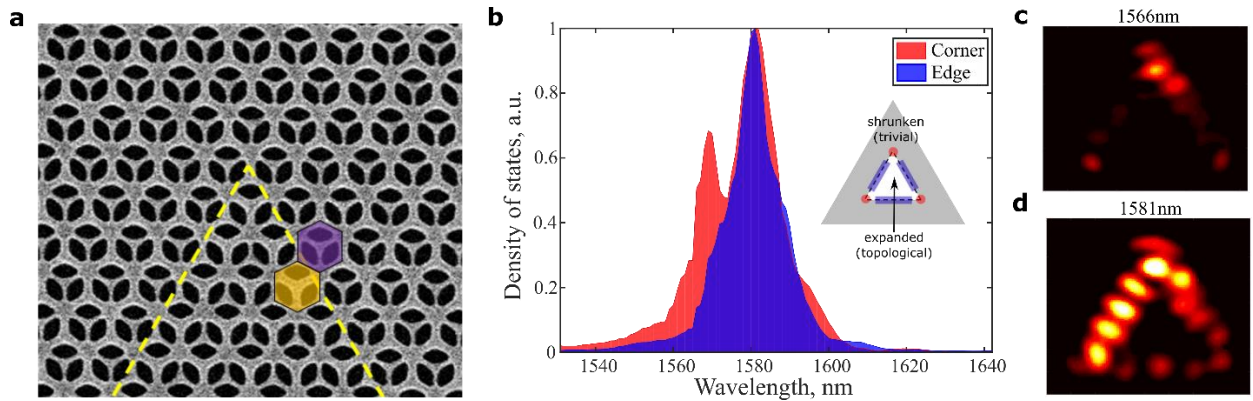

**Fig. 4| Experimental sample and results for imaging and density of states (DOS) in Kagome metasurface**. **a** SEM image of the sample showing the proximity of the corner of the topological (expanded) structure surrounded by the trivial (shrunken) lattice. Yellow dashed lines show the domain wall separating topological (yellow unit cell) and trivial (purple unit cell) domains. **b** Density of states extracted from far-field imaging at different wavelengths. A schematic diagram of metasurface is given in the inset. Corner and edge spatial filter-functions are marked by the corresponding colors, and the surrounding trivial region is indicated by the grey color. **c, d** Images of corner and edge states obtained by far-field imaging at the respective wavelengths.

In summary, we designed and experimentally demonstrated higher-order photonic topological metasurfaces supporting second-order corner states along with conventional (first-order) edge states. The topological transition associated with the opening of the Dirac cone below the light cone was experimentally revealed by using solid immersion ATR spectroscopy. Finally, dark-field imaging of the topological domain within the topologically trivial structure was performed to extract density of states and directly visualize topological edge and higher-order corner states supported by the metasurface. The presented demonstration of radiative topological states confined to one- and zero dimensions in the same structure opens a path towards novel applications of topological photonics to robustly control radiation and scattering of light. Future designs of metasurface HOTIs, where modes of different orders actively and controllably couple[29] with each other and with the radiative continuum, may lay the foundations of a novel platform for resilient control of radiation by trapping, guiding, scattering and radiating topological photons.

**Acknowledgements**

The work was supported by the Defense Advanced Research Projects Agency under the Nascent programme with grant number HR00111820040, by the Office of Naval Research, and by the National Science Foundation with grant numbers EFRI-1641069 and DMR-1809915. Research was partly carried out at the Center for Functional Nanomaterials, Brookhaven National Laboratory, which is supported by the U.S. Department of Energy, Office of Basic Energy Sciences, under Contract No. DE-SC0012704.

## Methods

### Simulations

For eigenmodes, band structures and Wannier center calculations, we used commercial software COMSOL Multiphysics 5.2a and 5.3 and its Radio Frequency module. For Wannier center calculation, the electrical field profiles under different **k** values needed for calculation are extracted by COMSOL Multiphysics with MATLAB from simulated kagome models. CST microwave studio software was used for the plane wave excitation calculations.

### Device fabrication.

Kagome lattice formed by the trimers of diamond shaped holes was fabricated on the Silicon-on-Insulator substrate (220 nm of Si, 3 µm of $SiO_2$) with the use of E-beam lithography (Elionix ELS-G100). First, the substrate was spin-coated with e-beam resist ZEP520A of approximately 300 nm thickness and then baked for 3 minutes at 180°C. E-beam lithography exposure was followed by the development process in n-Amyl Acetate cooled to 0C for 60 sec. Then cryogenic (-100C) anisotropic plasma etching of silicon was conducted in Oxford-C Plasmalab100 (etching rate of ~4nm per second). Diamond shaped holes were etched to the depth of about 220 nm. Finally, the residue of resist was removed by sample immersion into NMP for 4 hour.

Samples for solid immersion ATR spectroscopy represented bulk crystals of square shape with about 600 unit cells per side (~300µm), one is for ideal (unperturbed) Kagome and one is for expanded (perturbed) metasurfaces.

Far field imaging of topological states was performed on triangular shaped topological crystals surrounded by larger-sized trivial metasurface. Inner triangle – expanded crystal – had 20 unit cells per side (10.4µm) and cladding – shrunk crystal – had 140 unit cells per outer side (72.8 µm).

For some samples under-etching of silicon led to the spectral redshift of the modes, giving rise to variation in spectral positions of about 100nm from sample to sample, but we numerically confirmed that this had no effect on topological properties of the structures.

### Experimental characterization.

Schematic diagram of experimental setup for solid immersion ATR spectroscopy which allowed to recover topological transition is given in Add.Fig.1**a**. Light beam outcomes from the optical fiber with 400μm diameter which coupled with halogen light source. The beam was collimated by a lens with focal length 200mm and then focused up to the spot of diameter 200μm on the surface of the sample by a lens with focal length 100mm. The sample was attached on the rotational xyz-stage with both available axes of rotation – perpendicular to the plane of the sample and in-plane – that gave opportunity to vary $k_y$ and $k_x$, correspondingly. In order to observe modes under light cone we placed ZnSe hemispherical prism on top of the sample tuning the air gap between them by rotation the screw which pressed the prism. Reflected light was collected by optical fiber through a short-focus lens and then was guided to the Spectrometer (NIRQuest Ocean Optics). Total internal reflectance signal was measured for unattached ZnSe prism and was used as the reference for normalization of the spectrum received from the sample.

Direct visualization of edge and higher-order corner states was performed on the far field imaging setup (Add.Fig.1**b**). High-intensity supercontimuum light-source SuperK Extreme with connected LLTF Contrast tunable high-resolution bandpass filter generated light beam with 10nm bandwidth and the tunable wavelength in the range 0.9-2.0 µm. Then the light beam passed through a polarizer (getting linear p-polarization) and was focused by a lens ($f = 20$cm) on the tested metasurface under sharp angle of incidence ~70°. Implementing the dark-field mode the light scattered from the metasurface at the normal direction was collected by the reflective objective with NA=0.5 and 40x magnification. As back focal length of this objective was infinity, magnified image of topological crystal was directly captured by near-IR Xeva InGaAs camera, which was placed at approximately 1m away from the sample. Retrieving of density of states was performed through processing of a set of frames captured in the wavelength range 1.52-1.64 µm with 1nm resolution.

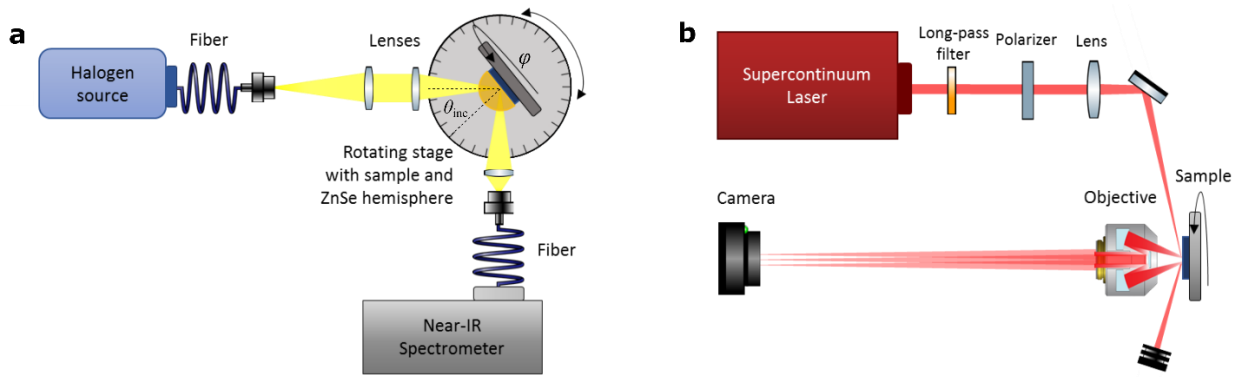

**Additional Fig.1|** Schematic diagram of experimental setups for solid immersion ATR spectroscopy (a) and far-field imaging (b).

**Data availability**

Data that are not already included in the paper and/or in the Supplementary Information are available on request from the authors.

**Author contributions**

All authors contributed extensively to the work presented in this paper.

**Competing interests**

The authors declare no competing interests.

**Corresponding authors**

Correspondence to Alexander B. Khanikaev and Andrea Alù.